\documentclass{aa}

\usepackage{graphicx}
\usepackage{subfigure}
\usepackage{txfonts}

\newcommand{\sax}{{\it BeppoSAX}}
\newcommand{\xmm}{{\it XMM-Newton}}
\newcommand{\sw}{{\it Swift}}
\newcommand{\sx}{{\it Swift--XRT~}}
\newcommand{\su}{{\it Swift--UVOT~}}
\newcommand{\ein}{{\it Einstein}}
\newcommand{\ros}{{\it ROSAT}}

\newcommand{\C}{3C~66A~}
\newcommand{\AO}{AO~0235+164~}
\newcommand{\SO}{S5~0716+714~}
\newcommand{\PK}{PKS~0823-223~}
\newcommand{\ON}{ON~231~}
\newcommand{\PKS}{PKS 0521-365~}

\newcommand{\OJ}{OJ 287~}

\begin{document}

\title{\sw~observations of IBL and LBL objects}

\author{
      F.~Massaro\inst{1,2}
     \and P.~Giommi\inst{3}
     \and G.~Tosti\inst{4}
     \and A.~Cassetti\inst{5}
     \and R.~Nesci\inst{5}
     \and M.~Perri\inst{3}
     \and D.~Burrows\inst{6}
     \and N.~Gerehls\inst{7}
}

\institute{Dipartimento di Fisica, Universit\`a di Roma Tor Vergata, Via della Ricerca scientifica 1 , I-00133 Roma, Italy
\and Harvard, Smithsonian Astrophysical Observatory, 60 Garden Street, Cambridge, MA 92138
\and ASI Science Data Center, ESRIN, I-00044 Frascati, Italy
\and Dipartimento di Fisica, Universit\'a di Perugia, Viale A. Pascoli 1, I-06123 Perugia, Italy
\and Dipartimento di Fisica, Universit\`a di Roma La Sapienza, Piazzale A. Moro 2, I-00185 Roma, Italy 
\and Department of Astronomy and Astrophysics, Pennsylvania State University, 525 Davey Lab, University Park, PA 16802
\and NASA NASA Goddard Space Flight Center, Greenbelt, MD 20771
}

\offprints{fmassaro@cfa.harvard.edu}
\date{Received ....; accepted ....}

\markboth{F. Massaro et al.:bla}
{F. Massaro et al.: bla.}

\abstract
{
BL Lacs are an enigmatic class of active galactic nuclei (AGNs), characterized by the non-thermal continuum typically attributed to synchrotron and inverse Compton emission. 
Depending on the frequency location of the maxima of these components, they are subdivided into three subclasses LBLs, IBLs, and HBLs.
We present the results of a set of observations of eight BL Lac objects of LBL and IBL type performed by the $XRT$ and $UVOT$ detectors onboard the \sw~satellite between January 2005 and November 2006. 
}
{
We are mainly interested in measuring the spectral parameters, and particularly 
the steepness between the UV and the X-ray band, useful for determining the classification of these sources.
We compare the behavior of these sources with previous \xmm, \sax~ observations and with historical data in the X-ray and in the optical band.
We are also interested in classifying the sources in our sample on the basis of the \sw~observations and comparing them with their classification presented in literature.
  }				
{
We performed X-ray spectral analysis of observed BL Lac objects using a simple powerlaw and in a few cases the log-parabolic model.
We also combined the UV emission with the low energy X-ray data to describe their spectral energy distribution.
}				
{
We used \sw~observational data to classify sources in our sample and derived parameters of their spectral energy distribution.
}				
{
We found that for the IBLs X-rays low states show features of the high energy component, usually interpreted as due to inverse Compton emission.
Sources in our sample exhibit a range of temporal UV and X-ray behaviors, some objects having clear and neat correlated UV and X-ray variations and other objects showing no clear  UV and X-ray correlation.
Finally, we also note that our estimates of spectral curvature are in the range of that measured for HBLs.
}	
			

\authorrunning{P. Giommi et al.}
\titlerunning{\sw~observations of IBL and LBL objects}

\maketitle

\section{Introduction}
In the unification scenario of active galactic nuclei (AGNs), BL Lac objects are interpreted as radio-loud sources with a relativistic jet that points toward us (Blandford \& Rees, 1978, Urry \& Padovani, 1995).
The BL Lac emission extends from radio to TeV energies and their spectral energy distribution (SED) is double-bumped: 
the first component typically peaks from the IR to X-ray band and the second one in the gamma-rays up to TeV energies.
Usually their SEDs are described in terms of Synchrotron Self-Compton (SSC) models in which synchrotron photons, emitted by a 
population of electrons accelerated in their relativistic jets, produce the second component via inverse Compton (IC) scattering with the same electrons. 
Padovani \& Giommi (1995) suggested the SED peak energy position of the first component as a possible classification criterium for BL Lacs,
indicating sources in which the synchrotron peak is between the UV band and X-rays as high-frequency peaked BL Lacs (HBLs)
and low-frequency peaked BL Lacs (LBLs) when the first bump appears in the IR-optical range.

Usually, X-ray observations of HBLs, up to $\sim$ 10 keV,  sample the maximum of the synchrotron component, while they only cover the spectral rise to maximum of inverse Compton component in LBLs.
In particular, there are some BL Lacs, named intermediate BL Lacs (IBLs) (see also Bondi et al. 2001), 
that present features of the two components in the 0.5-10 keV range. 
The \sax~observations of \ON had shown for the first time that this source could be considered a BL Lac object intermediate between the HBL and LBL classes, and 
this strengthened the relevance of the 0.1-10 keV energy range classifying the IBL sources.

It is well known that the optical spectrum of LBL and IBL objects generally becomes bluer when the source brightens this implies that the tail of the Synchrotron component moves to higher energies and may overcome the IC radiation. 
The classification of a given source as LBL or IBL may therefore depend on the luminosity state of the source at the epoch of observation. Actually the classification criteria are rather fuzzy. 
An attempt to determine the peak position of the SED for a large sample of sources in the northern hemisphere has been 
made by Nieppola, Tornikorsky and Valtaoja (2006) on the basis of non-simultaneus radio, optical, X-ray fluxes.
To further explore this topic, we selected a number of BL Lac objects from the Swift archive, classified in the literature as LBL or IBL, 
observed between January 2005 and November 2006 and not yet published, to search for signatures of the synchrotron tail and of the inverse Compton rise in the UV to 
the X-ray part of their spectrum.

In this paper, we present the \sw~data and we compare them with previous \xmm~ and \sax~ observations.
The list of the sources in our sample is shown in Table 1, their classification is based on the data available in the Blazar catalog (http://www.asdc.asi.it/bzcat) (Massaro et al. 2007).

\section{Observations and data reduction}
During our observations of IBLs and LBLs, the \sw~satellite was operated with all the instruments in data taking mode. 
We consider only $XRT$ (Burrows et al. 2005) and $UVOT$ (Roming et al. 2005) data, since our 
sources were not bright enough to be detected by the $BAT$ high energy experiment (Barthelmy et al. 2005).
In particular, \sx observations have been performed in photon-counting mode (PC) with only two exceptions: 
two pointings of \SO and one of \PKS, which were observed in the windowed-timing mode (WT) (see Sect. 2.1) discussed in the following.
Finally, we do not consider \sx observations with exposures shorter than 1000 seconds.
The log of \sx~and \su~observations is reported in Tab. 2.     
 
 \xmm~observations have been analyzed and discussed by several authors (see Foschini et al. 2006, and references therein).  
 Our analysis is in agreement with previous results. 
 \sax~and other historical data reported in this work are taken from the  the ASI Science Data Center (ASDC) archive.
  
The reduction procedure used to our investigations are described below.

\subsection{\sx}
The XRT data analysis has been performed with the XRTDAS software 
(v.~2.1), developed at the ASI Science Data Center (ASDC) and distributed within the HEAsoft package (v.~6.0.5). 
Event files were calibrated and cleaned 
with standard filtering criteria using the \textsc{xrtpipeline} task, combined with the latest calibration 
files available in the Swift CALDB distributed by HEASARC. 
Events in the energy range 0.3--10 keV with grades 0--12 (PC mode) and 0--2 (WT mode) were used in the analyses 
(see Hill et al.  2005 for a description of readout modes).

For the WT mode data, events for temporal and spectral analysis are selected using a 40 pixel wide 
(1 pixel corresponds to $2.36$ arcs) rectangular region centered on the source and aligned along the 
WT one dimensional stream in sky coordinates. 
Background events are extracted from a nearby source-free rectangular region of 40 pixel width. 

No signatures of pile-up were found in our \sx~observations.
Events are extracted using a 20 pixel radius circle. 
The background for PC mode is estimated from a nearby source-free circular region of 20 pixel radius. 

Ancillary response files are generated with the \textsc{xrtmkarf} task 
applying corrections for the PSF losses and CCD defects. 
The latest response matrices (v.~009) available in the Swift CALDB are used, and source spectra are binned 
to ensure a minimum of 30 counts per bin in order to utilize the $\chi^{2}$ minimization 
fitting technique and ensure the validity  of  Gaussian  statistics.

More details about the reduction procedure followed in this analysis is discussed in Massaro et al. (2008).

\subsection{\su}
The reduction procedure used to perform our \su~analysis is described in more detail in Tramacere et al. (2007a), 
we give only a brief description here. 

A variety of filter combinations and data modes are available for $UVOT$ 
observations.
For fields without bright stars, i.e. stars bright enough to degrade the 
detector, the most commonly-used  mode observes in six photometric bands: 
$U$, $B$, $V$, and three ultraviolet.  
In some cases our target was quite bright and saturated the image in all 
the photometric bands. 
For this reason not all the optical and UV data could be used for our analysis. 
The list of all available $UVOT$ observations is given in Table 2.

We performed the photometric analysis using a standard tool 
$UVOTSOURCE$ in HEAsoft package 6.0.5.
Counts were extracted from a 6$''$ radius aperture in the $V$, $B$,
and $U$ filters and from a 12$''$ radius aperture for the other UV filters
($UVW1$, $UVM2$, and $UVW2$), to properly take into account the wider PSF 
in these bandpasses.
The count rate was corrected for coincidence loss, and the background
subtraction was performed by estimating its level in an offset region at 
20$''$ from the source.
The estimate of flux uncertainties is complex due to possible instrumental 
systematics (e.g. residual pile-up in the central region of the PSF) and 
calibration, in particular, the lack of reference stars in the UV bandpasses. 
In this paper we adopt the conservative approach to consider a typical 
uncertainty of 8\% for the $V, B, U$ filters and of 15\% in the UV band.

The correction for the interstellar reddening was obtained assuming the 
$E(B-V)$ values at the source direction taken from NED and 
the fluxes were derived with the same conversion factors given by Giommi et 
al. (2006).

\subsection{\xmm}
The reduction procedure for \xmm~data is described completely in Tramacere et al. (2007b) and in Massaro et al. (2008).
Our sources were observed with \xmm~by means of
all EPIC CCD cameras: the EPIC-PN (Struder et al. 2001),  and EPIC-MOS  (Turner et al. 2001). 
We consider only EPIC-MOS data for our analyses.

Extractions of all light curves, source  and background spectra were done with
the \xmm~ Science  Analysis System (SAS) v6.5.0.  The Calibration
Index File  (CIF) and  the summary file  of the Observation  Data File
(ODF) were  generated using Updated Calibration Files (CCF) following the \"User's Guide  to
the \xmm~ Science Analysis System"  (issue 3.1, Loiseau et  al. 2004) and \"The \xmm~
ABC Guide"  (vers. 2.01, Snowden et al.  2004). 
Event files were produced by the \xmm~ EMCHAIN pipeline.

Light curves  for  each dataset are extracted,  and  all high-background intervals 
filtered out to exclude  time intervals contaminated  by  solar  flare signals. 
Then by direct inspection we select good time intervals far from solar flare peaks and 
with no count rate variations on time scales shorter than 500 seconds.
Photons are extracted from an annular region using
different apertures to minimize  pile-up, which affects MOS data.  The
mean value of the external radius for the annular  region is $40$~$''$.

A more restricted energy range (0.5--10 keV) is used to account
for possible residual calibration uncertainties.   
To ensure the validity  of  Gaussian  statistics,  data are  grouped  by  combining
instrumental channels so that each new bin comprises  40 counts or more.

\begin{table}
\caption{\sw~BL Lacs list}
\begin{flushleft}
\begin{tabular}{|lllll|}
\hline
Name & Class &    $RA$    &   $DEC$      & $N_{H,Gal}^*$    \\
\noalign{\smallskip}
\hline
\noalign{\smallskip}
0109+224         & IBL & 01 12 05.6  & +22 44 38.0  & 4.74             \\
 \C                       & IBL & 02 22 39.6  & +43 02 08.1  & 8.99             \\
 \AO                    & LBL &02 38 38.9  & +16 36 59    & 8.95             \\
\PKS                   & LBL &05 22 57.9  & -36 27 30.8  & 3.33             \\
\SO                     & IBL &07 21 53.4  & +71 20 36    & 3.81             \\ 
\PK                     & IBL & 08 26 01.6  & -22 30 27    & 8.51             \\
\OJ                     & LBL &08 54 48.9  & +20 06 30.9  & 3.04             \\
\ON                    & IBL &12 21 31.7  & +28 13 59    & 1.88             \\  
\noalign{\smallskip}
\hline
\end{tabular}
\end{flushleft}
(*) unit of $10^{20}cm^{-2}$ (Dickey \& Lockman, 1990)
\end{table}

\section{Spectral analysis}
For all sources the X-rays spectral fitting analysis was  performed with  the {\sc xspec} software package, version 11.3 (Arnaud 1996), 
with the galactic column densities fixed to the Galactic values (Dickey \& Lockman, 1990) and a single power-law model,
\begin{equation}
  N(E) = N_0 ~ \left( \frac{E}{1~ keV} \right)^{-\Gamma}~,
\end{equation}
where $\Gamma$ is the photon index.
We report in Table 2 the best fit values of $\Gamma$ and $N_0$, and the $\chi^2$ value of the fit.
the fit is good in a fair number of cases and its physical significance is discussed below for each source separately.

In IBLs, we search for possible signatures of both synchrotron and inverse Compton components by a direct inspection of residuals at the edges of 
the considered energy range.
We also try to describe their spectra with two more models: i) a broken power law and ii) a sum of two power laws, 
always with the absorption fixed to the Galactic value, to take into account of the possible presence of two spectral components. 
No improvement in $\chi^2$ was found using these extra two models for IBL sources.

More recently Tanihata et al. (2004), Massaro et al. (2004), and Tramacere et al. (2007b) have confirmed that a log-parabolic distribution 
gives very good fits for the synchrotron component of BL Lac sources,
\begin{equation}
  N(E) = N_0 ~ \left( \frac{E}{1~keV} \right)^{-a-b\log \left(\frac{E}{1~keV}\right)}
\end{equation}
or in the equivalent form and in terms of frequency:
\begin{equation}
  \nu F_{\nu} = S_{\nu} = S_p ~ \left( \frac{\nu}{\nu_p} \right)^{-b\log \left( \frac{\nu}{\nu_p} \right)}
\end{equation}
where $E=h\nu$.
This model provides the spectral description in terms of three parameters: the curvature, $b$; the SED peak frequency, $\nu_p$; 
and the height of the spectral energy distribution evaluated at $\nu_p$, $S_p=\nu F_{\nu}(\nu_p)$.
In the X-ray spectra of \SO and \ON, we tried to evaluate the presence of a high energy tail due to the Compton rise in the X-ray spectra using a log-parabolic spectral description  of the optical-to-soft-X-ray spectrum
(Landau et al. 1986, Massaro et al. 2004, Tramacere et al. 2007b). 

Furthermore, in analogy with \SO and \ON, we tried to describe the SED of 0109+224 and \C in terms of a log-parabolic model joining the UV 
data and the low-energy (less than 2 keV) X-rays data, deriving an estimate of the spectral parameters, even if the derived parameters have limited physical meaning due to the small amount 
of available data. We did not attempt to apply this description to \PK and \AO because the first source has no UV data available, while in the second one the
UV emission seems to belong to the tail of the synchrotron emission rather than being close to the peak of its SED (see Fig. 3 and Sect. 3.3 for details).

Details of single sources are discussed below.

\begin{table*}[!ht]
\caption{\sw Blazars Log Observations}
\begin{flushleft}
\begin{tabular}{|llllccccc|}
\hline 
$Obs ID$ &   Date           & UVOT   & XRT    & XRT  & $\Gamma$ & $N(E_1)$                                                      & $F_{2-10~keV}$                           &$\chi^2_{r}~(d.o.f.)$\\
                 &   dd-mm-yy   & Flters   & Frame & Exp   &                      & $10^{-4} ph~cm^{-2}~s^{-1}~keV^{-1}$  & $10^{-12} erg~cm^{-2}~s^{-1}$ &            \\
\hline   
\noalign{\smallskip}
\textbf{0109+224}&     &        &            &      &          &                  &        &            \\
00035001001 & 27-01-06 & all         & pc   & 1994 & ----         &  ----                 &0.56& ---- (3)   \\
00035001002 & 31-05-06 & all         & pc   & 7572 &2.16(0.09)&5.01(0.48)&1.02& 2.49 (12)  \\
00035001003 & 01-06-06 & V-B-U   & pc   & 19304&2.07(0.05)&4.99(0.31)&1.15& 0.96 (34)  \\ 
\hline  
\noalign{\smallskip}
\textbf{\C}&     &        &      &      &          &                  &        &            \\
00035003001 & 29-06-05 & all    & pc   & 5007 &2.34(0.08)&11.8(0.1)&1.83& 1.38 (16)  \\
00035003003 & 27-11-05 & all    & pc   & 51669&2.33(0.02)&14.9(0.3)&2.33& 0.97 (171) \\
\hline  
\noalign{\smallskip}
\textbf{\AO}&     &        &      &      &          &                  &        &            \\
00035004001 & 28-06-05 & no M2-W2    & pc   & 9710 &1.74(0.12)&3.27(0.62)&1.24& 1.44 (6)   \\
00035004002 & 07-07-05 & no B   & pc   & 11971&1.55(0.09)&3.18(0.44)&1.64& 1.50 (12)  \\
\hline  
\noalign{\smallskip}
\textbf{\PKS}&     &        &      &      &          &                  &        &            \\
00056640003 & 26-05-05 & W1     & pc   & 899  &1.64(0.10)&33.0(3.0)&14.8& 2.34 (10)  \\
00056640003 & 26-05-05 & W1     & wt   & 811  &1.93(0.09)&34.1(3.1)&9.70& 0.88 (11)  \\  
\hline  
\noalign{\smallskip}
\textbf{\SO}&     &        &      &      &          &                  &        &            \\
00030021001 & 02-04-05 & all    & pc   & 679  &  ----         &   ----                &1.28& ---- (3)   \\  
00030021001 & 02-04-05 & all    & wt   & 998  &   ----        &   ----                &2.07& ---- (6)   \\
00035009001 & 04-04-05 & all    & pc   & 7911 &2.67(0.06)&12.3(0.7)&1.21& 1.28 (31)  \\  
00035009001 & 04-04-05 & all    & wt   & 8969 &2.80(0.05)&14.5(0.7)&1.20& 1.50 (61)  \\
00035009002 & 18-08-05 & all    & pc   & 7414 &2.21(0.04)&24.7(1.1)&4.63& 1.10 (69)  \\ 
\hline  
\noalign{\smallskip}
\textbf{\PK}&     &        &      &      &          &                  &        &            \\
00035071001 & 20-06-05 & none   & pc   & 2897 &  ----         &  ----                 &0.56& ---- (2)   \\  
00035071002 & 01-07-05 & none   & pc   & 20288&2.10(0.07)&2.97(0.27)&0.65& 0.81 (16)  \\
00035071003 & 03-07-05 & none   & pc   & 31230&1.97(0.05)&2.78(0.22)&0.75& 0.91 (26)  \\
00035071004 & 06-09-05 & none   & pc   & 7345 &  ----         &   ----                &0.41& ---- (4)   \\    
00035071005 & 22-09-05 & none   & pc   & 5087 &  ----         &   ----                &0.48& ---- (3)   \\
00035071006 & 28-09-05 & none   & pc   & 5864 &  ----         &   ----                &0.49& ---- (4)   \\
00035071007 & 28-09-05 & none   & pc   & 8903 &  ----         &   ----                &0.52& ---- (7)   \\
00035071008 & 29-09-05 & none   & pc   & 16096&2.07(0.09)&2.56(0.31)&0.59& 0.57 (10)  \\
sum                   & Sep. 2005& none &  pc   & 67535 &2.16(0.07)&2.69(0.23)&0.55& 0.66 (26) \\
00035071009 & 02-10-05 & none   & pc   & 12346&  ----         &  ----                 &0.55& ---- (10)  \\
00035071010 & 01-12-05 & none   & pc   & 1935 &   ----        &  ----                 &   ----      & ----            \\    
00035071011 & 06-12-05 & none   & pc   & 9777 &   ----        &   ----                &0.64& ---- (6)   \\
\hline  
\noalign{\smallskip}
\textbf{\OJ}&     &        &      &      &           &                  &        &            \\
00035011001 & 20-05-05 & no B-U    & pc   & 3701 & ----          & ----                  &2.22& ---- (7)   \\
00035011002 & 26-05-05 & no U        & --   & -----&   ----        &  ----                 &   ----      &  ----           \\
00035011003 & 07-06-05 & none       & pc   & 1249 &   ----        &   ----                &2.68& ---- (5)   \\
00035905001 & 16-11-06 & none       & pc   & 1985 &   ----        &   ----                &1.42& ---- (2)   \\  
00035905002 & 17-11-06 & none       & pc   & 2255 &   ----        &   ----                &2.67& ---- (3)   \\ 
00035905003 & 18-11-06 & none       & pc   & 3147 &  ----         &   ----                &2.08& ---- (5)   \\ 
\hline  
\noalign{\smallskip}
\textbf{\ON}&     &        &      &      &          &                  &        &            \\
00035018001 & 14-07-05 & all       & pc   & 1398 &  ----        & ----                  &0.57& ---- (3)   \\
00035018002 & 29-10-05 & no B   & pc   & 9862 &2.67(0.08)&5.23(0.43)&0.52& 1.80 (17)  \\
00035018003 & 16-12-05 & all       & pc   & 8573 &2.65(0.06)&8.16(0.54)&0.89& 0.99 (25)  \\
\noalign{\smallskip}
\hline
\end{tabular}
\end{flushleft}
\end{table*}

\begin{table*}
\caption{\sx analysis with the log-parabolic model perfomed on \SO and \ON.}
\begin{flushleft}
\begin{tabular}{|lcccccc|}
\hline
\noalign{\smallskip} 
$Date$        & Frame & $a$ & $b$ &  $N(E_1)$       & $F_{2-10~keV}$   &$\chi^2_{r}~(d.o.f.)$\\
dd-mm-yy   &              &         &         & $10^{-4} ph~cm^{-2}~s^{-1}~keV^{-1}$  & $10^{-12} erg~cm^{-2}~s^{-1}$ &            \\
\hline
\noalign{\smallskip} 
\textbf{\SO}&&           &            &                   &          &           \\
00035009001 & wt & 2.79(0.04) & 0.76(0.13) & 13.1(0.8)& 2.34 & 1.08 (60) \\
\hline
\noalign{\smallskip}  
\textbf{\ON}&&           &            &                   &          &            \\
00035018002 & pc & 2.66(0.07) & 0.88(0.24) & 4.53(0.54)& 1.15 & 1.32 (16)  \\
\noalign{\smallskip} 
\hline
\end{tabular}
\end{flushleft}
\end{table*}

\subsection{0109+224}
The first X-ray detection of this source was performed by the IPC onboard \ein~between April 1979 and June 1980 (Owen et al. 1981). 
\xmm~and \sax~ did not observe of this source.

\sw~performed three observations of this IBL, but only on two occasions (2006 May 31 and 2006 June 1) and
the \sx exposure was larger than 5 ks, which is necessary to have a good statistics for a detailed spectral analysis.
The X-ray flux doesn't show significant variations during these observation. On the 2006 June 1 pointing it is about 20\% lower than the 2006 May 31 one.
The spectral shape is described well by a simple absorbed power law, with column density fixed to the Galactic value as reported in Table 2, 
and the X-ray photon index is close to 2, indicating that the SED is very flat.
Following the SSC scenario, this fact led us to conclude that X-ray data trace the IC component.

During the 2006 January 27 observation, Swift detected a low state
of the source in the UV band (5 times lower than the other two observations),
it is worth noting that this low state in UV corresponds to a not much lower state in the X-ray.

We evaluated the curvature, the peak frequency, and the height at the peak energy of the SED, as reported in Fig. 1, 
attributing the low energy X-ray emission to the tail of the synchrotron component. 
We found a curvature $b=0.19$, a peak frequency $\nu_p=3.17\times 10^{14}$~Hz, and the SED peak height $S_p=2.43\times 10^{-11}$$~erg~cm^{-2}~s^{-1}$ 
for the observation performed on 2006 May 31, and very similar values in the pointing of 2006 June 1: 
$b=0.20$, $\nu_p=3.43\times 10^{14}$~Hz, and $S_p=2.41\times 10^{-11}$$~erg~cm^{-2}~s^{-1}$, 
while on 27 January 2006 the spectral parameters are : $b=0.04$, $\nu_p=1.55\times 10^{11}$~Hz, and $S_p=2.94\times 10^{-11}$$~erg~cm^{-2}~s^{-1}$.
Such a low value of the peak frequency is physically unlikely:
the 2006 January 27 observation can be better explained if we assume that the X-ray flux is due to the sum of an IC emission, which remained nearly steady, and form the 
synchrotron emission that decreased significantly, as indicated by the UV flux.
A similar variability episode, with a steady IC and a large change in the synchrotron tail, was shown e.g. by BL Lacertae (Ravasio et al. 2002).
Considering the spectral shape, this source appears to be an IBL.

\begin{figure}[!htp]
\includegraphics[height=9cm,width=9.5cm,angle=0]{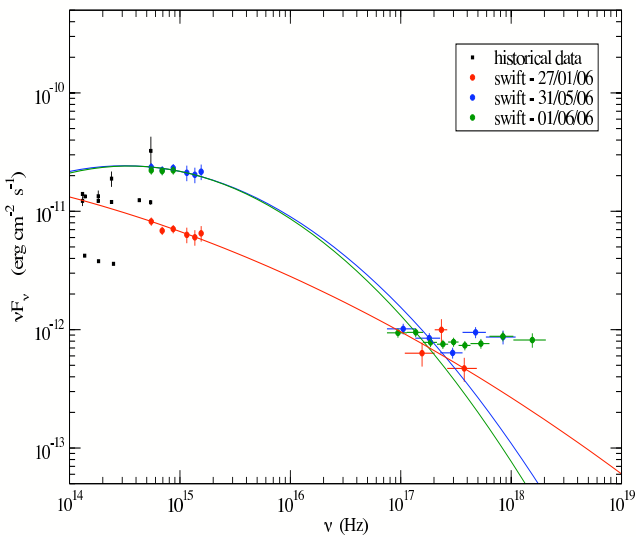}
\caption{The spectral energy distribution for the three observations of 0109+224 performed by \sw~compared to the historical data. The continuous curves reported in the plot refer 
to a log-parabolic model (see Sect. 3.1 for details).}
\end{figure}


\subsection{\C}
The source \C was extensively monitored from radio to very high energy $\gamma$-rays during 2003-2004 (B\"ottcher et al. 2005), and no significant X-ray variability was detected. 
Our analysis of the \sx~ and the \xmm~ data sets was performed with a simple, absorbed power law model with the column density fixed at the Galactic value, as reported in Table 1.
Our results agree with those reported by Croston et al (2003).
 
In detail, \sx performed two observations of \C during 2005. 
The photon index in the X-ray band is higher than 2 in each observation, 
indicating that X-ray emission is probably associated with the first bump in the SED, corresponding to the synchrotron radiation in an SSC scenario.
 
In particular, this BL Lac was found in two differents X-ray states. 
In the low state (2005 June 29), the source seemed to show an excess,  in the residuals at the edge of the X-ray band, when fitted with a simple power law model.
It is probably due to the start of the Compton rise of the second bump, although the X-ray fit alone 
cannot distinguish bewteen a simple absorbed power-law or a multi-component model because the statistics in the range 5-10 KeV is very narrow.
The high state observation (2005 November 27) is described well by a power law with a flat spectral index in the SED.
Its spectral energy distribution is shown in Fig. (2).

The ratio between the X-ray flux of the two \sx observations is about 1.57.
It is worth noting that the high UV state of the source corresponds to a low state in X-rays, in agreement with the possibility that the low energy X-ray emission corresponds to the synchrotron tail. 
Spectral parameters evaluated for the observation performed on 2005 June 29 are: 
$b=0.15$, $\nu_p=3.41\times 10^{14}$~Hz, and $S_p=2.78\times 10^{-11}$$~erg~cm^{-2}~s^{-1}$, 
while in the observation performed on 2005 November 27: $b=0.16$, $\nu_p=4.78\times 10^{14}~Hz$, 
and $S_p=2.20\times 10^{-11}$$~erg~cm^{-2}~s^{-1}$, showing a large spectral change in the SED peak frequency.  
\xmm~data, reported in Fig. (2), agree with the first \sx observation showing the same feature.
Comparing the \sx data with the historical ones (\ein, \ros, \sax), we see that the Compton rise seem to always appear in low states, as just discussed by Perri et al. (2003).

\begin{figure}[!htp]
\includegraphics[height=9cm,width=9.5cm,angle=0]{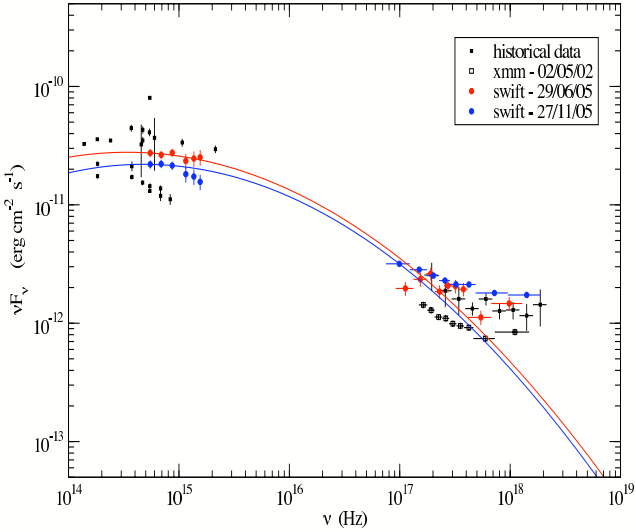}
\caption{The spectral energy distribution for the two observations of \C performed by \sw, in comparison with one \xmm~pointing and with historical data. The continuous curves reported in the plot refer to a log-parabolic model (see Sect. 3.2 for details).}
\end{figure}


\subsection{\AO}
\AO is one of most variable sources in the sky, it usually displays strong variability from radio to optical wavelengths.
It has a very compact nucleus, and sometimes jet features with high superluminal motion can be observed (Jorstad et al. 2001).
We found that \xmm~ observations are strongly affected by pile-up that can affect the estimate of the spectral index, even if the source is in a very low energy state, as reported in Fig. 3.   
Finally, as shown in the historical SED plotted in Fig. 3, this source is also extremely variable in the X-ray band.

\sx observed \AO on two occasions, on 2005 June 28 and on 2005 July 7, finding the source in a very similar state. 
The X-ray ratio between the 0.2-10 keV fluxes of the two observations is 1.36.
The photon index estimated by the X-ray band best fit is lower than 2 indicating, that these emission could be due to the inverse Compton component.
Compared to \xmm, \sx found the source in an intermediate state consistent with other spectra observed by \sax~(Perri et al. 2003).
In particular, \su and \sx seems to belong to different components, the first to the tail of the synchrotron emission and the second to the rise of the inverse Compton bump.

From our analysis we argue that \AO appears to be a LBL object.
\begin{figure}[!htp]
\includegraphics[height=9cm,width=9.5cm,angle=0]{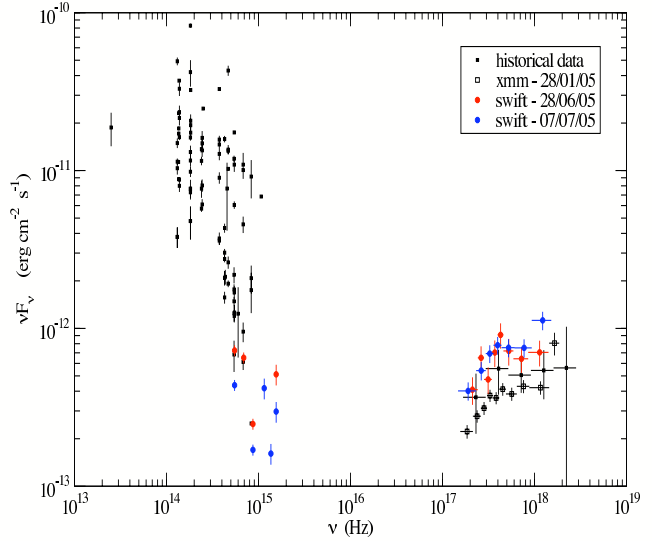}
\caption{The spectral energy distribution for the two observations of \AO performed by \sw, compared with one \xmm~pointing and with historical data.}
\end{figure}

\subsection{\PKS}
\PKS is a well-studied BL Lac with a redshift of 0.055, comparable to the redshifts of the radio galaxies
studied by Worrall \& Birkinshaw (1994). It is variously described in the literature as a blazar, a BL Lac object, or
an N-galaxy; and on multifrequency spectral index plots like those of Sambruna, Maraschi \& Urry (1996), it is placed among
radio-selected BL Lacs.
It has already been extensively observed at X-ray wavebands, with \ein~(Worrall \& Wilkes 1990), 
$EXOSAT$ (Sambruna et al. 1994), and the \ros~PSPC (Pian et al. 1996), with an X-ray flux at 1 
keV of about 1.2 $\mu$Jy, 1.7 $\mu$Jy, and 2.1$\mu$Jy, respectively. It was also detected in $\gamma$-rays 
by $EGRET$ (Lin et al. 1996). 
The \xmm~ observation, performed on the 2002 October 9, showed that the source appears to have a concave spectrum in the X-ray waveband with the photon index of 
about $2.0$ in the soft band, flattening to about $1.74$ at higher energies (Foschini et al. 2006).

The \sx observation found this source in a higher state compared to archival data. 
We analyzed the combined PC and WT observation to have enough useful counts estimating 
the photon index of the 2005 May 26 pointing. This was found about 1.86, thus confirming the LBL classification of \PKS indicating that the X-ray data belongs to the IC component.
The spectral energy distribution of \PKS is reported in Fig. 4.

\begin{figure}[!htp]
\includegraphics[height=9cm,width=9.5cm,angle=0]{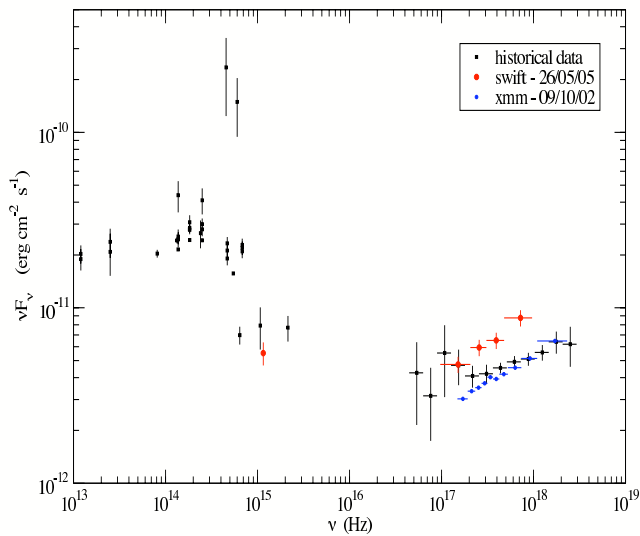}
\caption{The spectral energy distribution for the observation of \PKS performed by \sw, compared with historical data and with the \xmm~observation.}
\end{figure}

\subsection{\SO}
This is the most famous IBL source. 
The past X-ray observations, performed with \ros~(Cappi et al. 1994), have shown strong variations with short flares. 
Broad-band observations with \sax~were performed in 1996 and 1998 during low optical activity (Giommi et al. 1999) and in 2000 (Tagliaferri et al. 2003), following a
strong optical outburst, when the blazar displayed an increase of 2.3 magnitudes in 9 days (Raiteri et al. 2003).
In the first two observations, \SO was at a low flux level in the 0.1 - 10 keV energy band (3.4$\times 10^{-12} $$~erg~cm^{-2}~s^{-1}$ 
 in 1996 and 4.4$\times 10^{-12} $$~erg~cm^{-2}~s^{-1}$ in 1998) and was not detected at energies above 10 keV. 

The soft X-rays displayed significant variability, in agreement with the optical light curves, with the possibility that the optical
and X-ray variations were correlated, but the poor statistics prevented the authors from making firm conclusions (Giommi et al. 1999).
The \sax~observation of 2000 confirmed the findings by Giommi et al. (1999), according to which the soft X-ray spectrum becomes steeper when the source flux increases.
In particular, the \sax~spectral energy distribution showed the presence of a  steep power-law component below $\sim$ 2.5 keV with a photon index $\sim$ 1.5,
becoming harder towards higher frequencies, with a new spectral index $\sim$ 0.85 (Giommi et al. 1999).
The source was also detected in hard X-rays up to 60 keV.
Again, variability appears to be present only in the soft X-rays (Tagliaferri et al. 2003).

In April 2004, \xmm~(Foschini et al. 2006) and INTEGRAL (Pian et al. 2005) observed \SO nearly simultaneously, triggered by an optical outburst that occurred at the end of March. 
The X-ray spectrum is well-represented by a concave broken power-law model, with the break at about 2 keV. 
In the framework of the synchrotron self-Compton model, the softer part of the spectrum, which is described by a power law of index $\Gamma \simeq 1.8$, is probably due to synchrotron emission, while the
harder part of the spectrum, which has $\Gamma \simeq 1$, is due to inverse Compton emission. 

Compared to the \xmm~observation the \sx~pointings are in lower states, which may help detecting features of the rise in the Compton emission as shown in the \sax~observation reported in Fig. 6.
In fact, in the \sx observation on 2005 April 4 with a log-parabolic model, we found a positive value of the curvature parameter, (Table 3) only for the X-ray band, consistent with the presence of a flattening of the spectrum toward high energies.
Spectral parameters of the 2005 April 4 and2005 August 18 observations, evaluated linking the 
UV data with the low energy X-ray emission, are $b=0.22$, $\nu_p=6.36\times 10^{14}$, 
$S_p=4.74\times 10^{-11}$, and $b=0.20$, $\nu_p=4.11\times 10^{14}$ and $S_p=9.87\times 10^{-11}$.

The spectral energy distribution is reported in Fig. 6, where it is possible to see the deviations in the tail of X-ray emission with respect to the log-parabolic model, within all observations.

\begin{figure}[!htp]
\includegraphics[height=9cm,width=9.5cm,angle=0]{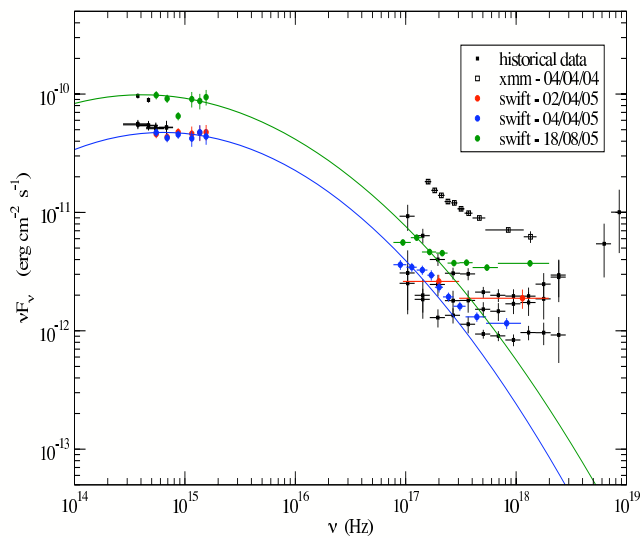}
\caption{The spectral energy distribution for the three observations of \SO performed by \sw, in comparison with one \xmm~pointing and with historical data. The continuous curves reported in the plot refer to a log-parabolic model (see Sect. 3.5 for details).}
\end{figure}


\subsection{\PK}
This BL Lac object (z=0.91) was never detected in X-rays prior to \sx.
The X-ray flux measured in the \xmm Slew Survey is consistent with zero within 3 sigma range.
The X-ray photon index in the long exposure observation is very close to 2, indicating a flat spectral energy distribution.
The 0.2-10 keV flux increases by a factor of 1.16 between the pointing performed on 2005 July 1 and the 2005 July 3 observation,
and it decreases by a factor of 0.73 from July to September 2005.
We report in Table 2 the sum of all observations performed in September 2005 in order to increase statistics since during
these observations there are no large flux variations. The previous classification of this source was based on the flat SED in the optical band (JHK), similar to \C.
Our new \sx observations confirm that this source belongs to the IBL class.


\subsection{\OJ}
The BL Lac object \OJ (z = 0.306; Stickel et al. 1989) is one of the well observed LBL sources.
In particular, it was the main target of an intensive optical (Pursimo et al. 2000) and radio (Valtaoja et al. 2000) monitoring.
One of its most relevant characteristics is the periodicity of 11.95 years, derived from a historical optical light curve (Sillanp\"a\"a et al. 1988).
\sax~observations (Massaro et al. 2003) report that the simple power law extrapolations of the optical data into the X-ray range are higher than the measured fluxes.
This implies that the spectral index must steepen significantly toward the UV Ð X range, unless the emission originates in different components.

We argue that the optical points belong to the synchrotron component, whereas the measured fluxes in the X-ray range 
are on the Compton branch and represent only upper limits for the high frequency synchrotron tail. 
In the case of this source, the synchrotron component is not detected in the X-ray range and the source can be classifed as LBL;
the estimated position of the peak frequency is at about 2.2 $\times 10^{13}$~Hz (Massaro et al. 2003).

\sw~observed \OJ three times during 2005 and on other three occasions in 2006. During 2005, the X-ray flux is similar in each pointing, while in 2006 
the ratio between the two observations performed on 2006 November 17 and 2006 November 16  is about 1.88. Short exposures in \sx pointings did 
not permit us to evaluate spectral parameters.
The average flux in the range $2-10~keV$ among 2005 \sw~pointings is $2.45\times 10^{-12}$$~erg~cm^{-2}~s^{-1}$ and $2.06\times 10^{-12}$$~erg~cm^{-2}~s^{-1}$

The comparison with \xmm~ observation performed on April 2005 is reported in Fig. 7, and other \xmm~data available are consistent with them. 
The flux estimated with \sx~is consistent with that of \xmm~ (Ciprini et al. 2007a), but the low statistics due to the short exposure observations \sw~does not allow us 
to look for a possible energy break in the \sx~spectra of \OJ (Ciprini et al. 2007b; Ciprini et al. 2008, in prep.). 
As shown in Fig. 6, comparing \sx~observatons with historical data no significant changes in the SED are evident.

\begin{figure}[!htp]
\includegraphics[height=9cm,width=9.5cm,angle=0]{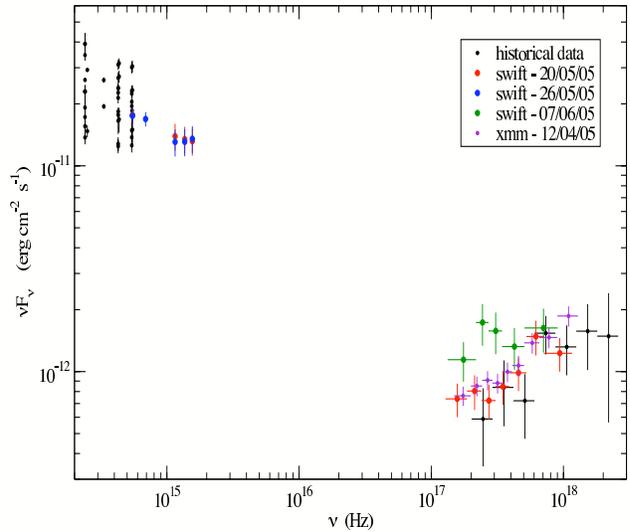}
\caption{The spectral energy distribution for the 2005 observations of \OJ performed by \sw, compared with one \xmm~pointing and with historical data.}
\end{figure}

\subsection{\ON}
The BL lac object \ON (1219+285, W Comae, B2 1219+28, $z=0.102$) was observed in the 
X-ray, for the first time by $Einstein$ IPC (Giacconi et al. 1979) in June 1980. The
\sax~observations (Tagliaferri et al. 2000) measured the hard X-ray spectrum above $3$ keV and in different brightness states for the first time . 
The \sax~spectral analysis, performed with a broken-powerlaw and a sum of two power law models, had shown a concave shape for the spectrum in X-ray band, indicating a first photon index that was very similar to the previous X-rays \ein~observations, with an energy break at lower energies ($4.00$ keV and $2.60$ keV for the two different \sax~observations), and a second photon index steeper than the previous one.
The shape of the spectrum determined by \sax~was found to be different from \ros, described as a single power law with a photon index 1.19 (Comastri et al. 1997) , and this was interpreted as due to the narrower energy range of the different instruments and a weaker state of \ON itself (Tagliaferri et al. 2000).
Finally, the concave shape of the spectrum was interpreted in terms of presence of the steep tail of synchrotron radiation (first power law) and 
the hard emerging of the inverse Compton emission (second power law).

The photon index of \sx observation is larger than 2, indicating that X-ray spectra belong to the tail of the synchrotron emission. 
The first \sx observation (2005 October 29) confirms the previous results showing both components in the low state, a similar feature to other IBL objects.
The lower statistics in the high energy tail does not permit evaluating of any spectral index in the Compton rise.
The second \sx observation (2005 December 16) showed the source in a state a bit fainter than the XMM-Newton one: the spectral slope is very similar and no clear evidence of upturn of the SED at high energies is detectable
This effect is due to the narrow energy range of \xmm~ and \sw; in fact, in the historical data (\sax~ observation), representing a high state of this source, the inverse Compton rise can be seen, due to the wider energy range PDS instrument. While considering only the low energy camera below 10 keV, this spectrum is described well in terms of a single power law. 

The 0.2-10 keV flux increases 80\% between the two states observed by \sw.  
Spectral parameters evaluated connecting UV data with low energy X-ray emission are: $b=0.17$, $\nu_p=2.45\times 10^{14}$~Hz, $S_p=1.52\times 10^{-11}$$~erg~cm^{-2}~s^{-1}$, $b=0.25$, $\nu_p=6.23\times 10^{14}$~Hz, $S_p=2.69\times 10^{-11}$$~erg~cm^{-2}~s^{-1}$ and $b=0.19$, $\nu_p=4.10\times 10^{14}$~Hz and $S_p=3.30\times 10^{-11}$$~erg~cm^{-2}~s^{-1}$ for the three \sx pointings performed on 2005 July 14, 2005 October 29, and 2005 December 16.  
The spectral energy distribution of \ON is shown in Fig. 7. 

\begin{figure}[!htp]
\includegraphics[height=9cm,width=9.5cm,angle=0]{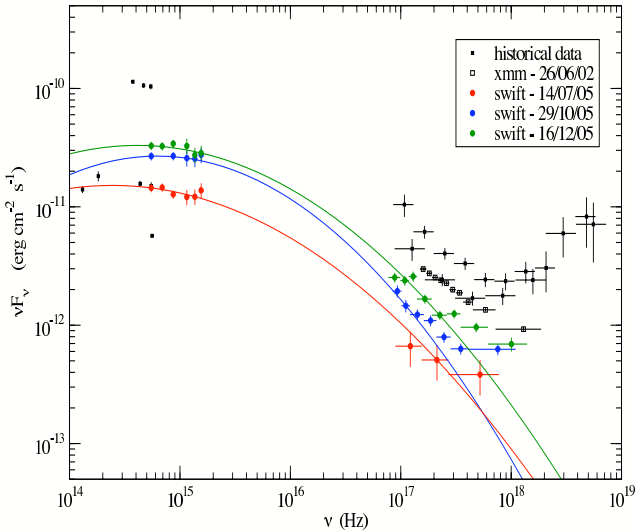}
\caption{The spectral energy distribution for the 2005 observations of \ON performed by \sw, in comparison with one \xmm~pointing and with historical data. The continuous curves reported in the plot refer to a log-parabolic model (see Sect. 3.8 for details).}
\end{figure}


\section{Discussion}
In the present paper we have described the results of the observations of eight BL Lac objects performed with the
$XRT$ and $UVOT$ detectors onboard the \sw~satellite between January 2005 and November 2006.
Among these sources, the first X-ray detection of \PK is reported.
The X-ray observations combined with UV data are crucial classifying BL Lacs and understanding the physical processes
affecting both synchrotron and inverse Compton components.

Our data show that the SEDs of IBL objects in low brightness states are characterized by a very apparent inverse Compton
rising component in the high energy tail, while the UV emission, combined with the soft X-ray emission belongs to the
synchrotron component. In some cases we were able to estimate the SED peak frequencies and the spectral curvatures.
We stress that these values agree closely with those that can be derived from the best fit parameters of the LBL
and IBL sample considered by Landau et al. (1986), for which curvatures were in the range from
0.09 to 0.22 with a mean value of 0.14.

This finding suggests that the SEDs of the synchrotron component have remarkably similar shapes, most likely because
the physical conditions and acceleration mechanism are basically the same.
Similar values of the curvature parameter $b$ are also found in the HBL sources observed in bright states
with \sax~(Giommi et al. 2005), or with \xmm~and \sw~as reported in Tramacere et al. (2007a) and Massaro et al. (2008).
For the few LBLs of our sample, the single power law model provides a good description of the X-ray spectra, supporting the idea that they are basically produced by an Inverse Compton process.

The reason for the subdivision of BL Lac objects into the two main types of LBL and HBL is not yet fully understood, 
and therefore the study of transitions objects, like IBL sources, is important to study the differences between them.
These differences can be due to either the relativistic beaming or to other physical parameters as the acceleration efficiency.

For the well known source S5 0716+714, for instance, Nesci et al. (2005) on the basis of the historic light curve and VLBI data on the evolution of superluminal motion (Bach et al. 2005),  proposing that the observed changes are due to a slow increase in the beaming factor, likely caused by the jet direction
approaching the line of sight.

To verify this hypothesis, and to gain more confidence on a substantial similarity of LBL and IBL sources,
an independent estimate of the physical parameters  will be very useful . 
Under this respect simultaneous observations of both synchrotron and
inverse Compton components are very useful, in particular when associated with variability studies. 
This possibility will soon be offered by the upcoming space missions GLAST
in the $\gamma$-ray band, and Planck in the microwave-IR range, together
with ground-based and X-ray observation, that will contribute to much denser
coverage of the SEDs, and will particularly give the possibility of investigating
how the ratio between the synchrotron and inverse Compton peak frequencies
changes when the source luminosity brightens or dims.

\begin{acknowledgements}
We thank our referee E. Pian for helpful comments improving our presentation.
F. Massaro  thanks G. Cusumano for his help in the use of the \sx~ data reduction procedure, and 
A. Tramacere for his suggestions and discussions on the use of \xmm~SAS software. 
We thank S, Ciprini for helpful discussion of \OJ \xmm~observations.
This research has made use of the NASA/IPAC Extragalactic Database (NED), 
which is operated by the Jet Propulsion Laboratory, California Institute of Technology, 
under contract with the National Aeronautics and Space Administration.
\end{acknowledgements}


\end{document}